\documentclass[fleqn,usenatbib]{mnras}

\usepackage{newtxtext,newtxmath}

\usepackage[T1]{fontenc}
\usepackage{ae,aecompl}

\usepackage{graphicx}	
\usepackage{amsmath}	
\usepackage{amssymb}	

\title[Absorption-selected galaxies with evidence for excited ISMs]{ALMACAL V:  Absorption-selected galaxies with evidence for excited ISMs}

\author[A. Klitsch et al.]{
A. Klitsch,$^{1, 2}$\thanks{E-mail: aklitsch@eso.org}
M. A. Zwaan,$^{1}$ 
C. P\'eroux,$^{3}$ 
I. Smail,$^{2}$ 
I. Oteo,$^{1}$ 
G. Popping,$^{4}$ 
\newauthor{
A. M. Swinbank,$^{2}$
R. J. Ivison,$^{1}$
and A. D. Biggs$^{1}$ }\\
$^{1}$European Southern Observatory, Karl-Schwarzschild-Str. 2, 85748 Garching near Munich, Germany\\
$^{2}$Centre for Extragalactic Astronomy, Durham University, Department of Physics, South Road, Durham DH1 3LE, UK\\
$^{3}$Aix Marseille Univ, CNRS, LAM, (Laboratoire d'Astrophysique de Marseille), UMR 7326, 13388, Marseille, France\\
$^4$Max-Planck-Institut f\"ur Astronomie, K\"onigstuhl 17, 69117 Heidelberg, Germany}
\date{Accepted XXX. Received YYY; in original form ZZZ}

\pubyear{2018}

\begin{document}
\label{firstpage}
\pagerange{\pageref{firstpage}--\pageref{lastpage}}
\maketitle

\begin{abstract}
Gas-rich galaxies are selected efficiently via quasar absorption lines. Recently, a new perspective on such absorption-selected systems has opened up by studying the molecular gas content of absorber host galaxies using ALMA CO emission line observations. Here, we present an analysis of multiple CO transitions ($L'_{\rm CO} \sim 10^9$ K km s$^{-1}$) in two $z \sim 0.5$ galaxies associated with one Ly~$\alpha$ absorber towards J0238+1636. The CO spectral line energy distribution (CO SLED) of these galaxies appear distinct from that of typical star-forming galaxies at similar redshifts and is comparable with that of luminous infrared galaxies or AGN. 
Indeed, these galaxies are associated with optically identified AGN activity.
We infer that the CO line ratios and the $\alpha_{\rm CO}$ conversion factor differ from the Galactic values. Our findings suggest that at least a fraction of absorption selected systems shows ISM conditions deviating from those of normal star-forming galaxies. For a robust molecular gas mass calculation, it is therefore important to construct the CO SLED. 
Absorption-line-selection identifies systems with widely distributed gas, which may preferentially select interacting galaxies, which in turn will have more excited CO SLEDs than isolated galaxies. Furthermore, we raise the question whether quasar absorbers preferentially trace galaxy overdensities.

\end{abstract}




\section{Introduction}

The neutral phase of the gas traced by HI is thought to be the original reservoir for star formation (SF) \citep{Wolfe1986}. Quasar absorbers associated with foreground galaxies provide a unique tool to probe this neutral gas phase down to low gas densities and at any redshift. 
However, the physical processes that transform H{\sc i} into molecular gas and hence stars remain unconstrained. A direct probe of the fuel for SF has to come from measurements of molecular gas. Molecular Hydrogen (H$_2$) absorbers at UV wavelengths are only detected in $\approx 50$\% of the  H{\sc i}-rich systems at low redshift \citep{Muzahid2015} and $\approx 15$\% at high redshift \citep{Noterdaeme2008}. Furthermore, the median molecular fractions in absorbers with H$_2$ detections reported are low: $\log f({\rm H}_2)$ is $-1.93 \pm 0.63$ at low redshift and $-2.3 \pm 0.8$ at high redshift showing no significant evolution. 
These low detection rates in absorption are probably a consequence of the low sky cross section of molecular gas compared to H{\sc i} neutral gas \citep{Zwaan2006}. \\
\indent Progress in relating the properties of H{\sc i}-rich systems to their SF has to come from identifying the galaxy counterparts to quasar absorbers. Theoretical models suggest the true absorbers, at least those with large column density of neutral gas, are associated with galaxies 
\citep{Fumagalli2011,vandeVoort2012,Bird2014}. The observational challenge of identifying the absorber host galaxy in emission has been partly overcome using high-z absorbers to block the light from the background quasar in direct imaging \citep[e.g.][]{Fumagalli2010}, using slit triangulation \citep[e.g.][]{Krogager2017}, and  thanks to integral field units (IFU) \citep[e.g.][]{Peroux2011,Fumagalli2017,Peroux2017,Rudie2017,Klitsch2018}. 
Several studies report Ly$\alpha$ absorbers that are associated to galaxy groups instead of isolated galaxies \citep[e.g.][]{Bielby2016, Rahmani2017, Fumagalli2017, Peroux2017}. Furthermore, Ly$\alpha$ absorber host galaxies exhibiting strong winds are reported in the literature \citep[e.g.][]{Rudie2017, Fynbo2018}. Although the number of observations of quasar absorber host galaxies is rising, the impact of selection criteria (high metalicity, etc.) and different observation techniques is unclear. Therefore, the question of exactly what population of galaxies is traced by the absorption remains.

Recently, a new perspective on relating baryons traced by cold gas and stars opened up with reports of CO rotational transitions emission in quasar absorber galaxy hosts observed with ALMA.
To date ten molecular gas-rich absorption-selected galaxies with large inferred molecular gas masses of $10^{10} - 10^{11}$~M$_{\sun}$ have been analysed \citep{Neeleman2016,Neeleman2018,Kanekar2018, Mller2018, Klitsch2018, Fynbo2018}. 
The combination of low detection rates of H$_2$ absorption and the high molecular gas masses inferred from CO observations is a key question in this field.
Here we report the detection of two galaxies at the redshift of one Ly~$\alpha$ absorber in our \mbox{(sub-)}millimetre survey ALMACAL \citep{Oteo2016}. For each of these, we detect multiple CO transitions, which offers fresh clues on the molecular gas conditions of these galaxies.

\section{Sample and Reduction}

\begin{table}
\caption{Summary of the ALMA observations of J0238+1636}
\label{TabSummaryCubes}
\centering{
\begin{tabular}{c c l l l l}
\hline
Band & ang. res. & rms & vel. res & $t_{\rm int}$\\
  & [\arcsec] & [mJy bm$^{-1}$] & [km s$^{-1}$] & [ks]\\
\hline
 4 & 0.43 & 0.18 & 36 & 0.6\\
 6 & 0.96 & 0.24 & 25 & 5.0\\
 7 & 0.67 & 0.14 & 20 & 1.5\\
 7 cont. & 0.32 & 0.04 & -- & 5.1\\
\hline
\end{tabular}}
\end{table}

\begin{table*}
\caption{ALMA detections of multiple CO emission lines at the redshift of two absorbers towards J0238-1636 and J0423-0120}
\label{TabGalEssential}
\begin{tabular}{l l l l l r r c c c c}
\hline
Name & $z$ & R.A. & Dec. & $\theta$ & $S_{\rm int}^{\rm cont.}$ & CO & $S_{\rm peak}$ & $S_{\rm int}$ & $\log (L'_{\rm CO})$ & FWHM\\
 & & (J2000) & (J2000) & [kpc] & [mJy] & trans. & [mJy] & [Jy km s$^{-1}$] & [K km s$^{-1}$ pc$^{2}$]& [km s$^{-1}$]\\
\hline
 J0238A & 0.524 & 02 38 38.94 & +16 36 57.3 & $12$ & $<0.1^b$ & 2-1 & $1.5 \pm 0.2$ & $0.32 \pm 0.06$ & $9.07 \pm 0.06$ & $350 \pm 20$\\
 &  & & & & & 3-2 & $2.3 \pm 0.2$ & $0.70 \pm 0.09$ & $9.05 \pm 0.09$ & $380 \pm 40$\\
 & & & & & & 4-3 & $4.1\pm 0.9$ & $1.2 \pm 0.4$ & $9.0 \pm 0.4 $ & $420 \pm 30$\\
J0238A1 & 0.524 & 02 38 39.01 & +16 36 59.2 & $7$ & $<0.1^b$ & 2-1 & $0.7 \pm 0.2$ & $0.2 \pm 0.05$ & $8.87 \pm 0.05$ & $350 \pm 40$\\
 & & & & & & 3-2 & $1.4 \pm 0.2$ & $0.43 \pm 0.04$ & $8.8 \pm 0.1$ & $380 \pm 40$\\
  & & & & & & 4-3 & - & $<0.9$ & $<9 $ & - \\
 J0423B &0.633 & 04 23 16.07 & $-$01 20 52.1 & $133$ & $0.8 \pm 0.2^a$ & 2-1 & $8.4 \pm0.7$ & $3.6 \pm 0.2$ & $10.28 \pm 0.02$ & $590 \pm 30$\\
 && & & & & 3-2 & $10 \pm 1$ & $5.2 \pm 0.3$ & $10.10 \pm 0.02$ & $610 \pm 40 $\\
 \hline
\end{tabular}
\begin{minipage}{\linewidth}
$\theta$ denotes the impact parameter between the galaxy and the background quasar. The properties of J0423B are taken from \citet{Klitsch2018}. $^a$ denotes the Band 6 and $^b$ denoted the Band 7 continuum flux. The upper limit on the continuum flux for J0238A and J0238A1 are based on the $3 \sigma$ flux limit from the Band 7 map and assuming that they are unresolved. The upper limit of the CO(4-3) flux from J0238A1 is based on a $3\sigma$ flux limit assuming a FWHM of 350~km~s$^{-1}$.
\end{minipage}
\end{table*}

In this study we use ALMACAL, a \mbox{(sub-)}millimetre survey utilizing calibrator observations from the ALMA archive \citep{Oteo2016}. We search for CO emission lines from host galaxies of known quasar absorbers. Cross-correlating the sky positions of known absorbers detected at optical and UV wavelengths with the 749 calibrators in our ALMACAL catalogue leads to 109 positional matches. Matching the frequency coverage in ALMACAL with the redshift of the absorbers indicates a coverage of 57 absorbers in the redshift range $0.06 < z_{\rm abs} < 3.0$ towards 26 quasars for which the frequency of at least one CO transition is observed. Here we focus on absorber host galaxies with multiple CO emission line detections in order to study the CO excitation. We discover two new molecular gas-rich galaxies with multiple CO emission lines detected at the redshift of an H{\sc i} absorber towards the quasar J0238$+$1636 at $z_{\rm abs} = 0.524$. Additionally, we include the molecular gas-rich galaxy detected in the field of J0423$-$0120 ($z_{\rm abs} = 0.633$) \citep{Klitsch2018} in our analysis.

The retrieval of the calibrator data is described in detail in \cite{Oteo2016}; the calibration and data reduction are carried out as described in \citet{Klitsch2018}. The properties of the final data cubes are summarised in Table~\ref{TabSummaryCubes}.

The gas properties of the absorbers, CO detected host galaxies and impact parameters are given in Table~\ref{TabGalEssential} and \ref{TabPhysProp}, taken from \citet{Klitsch2018, Rao2006, Junkkarinen2004, Burbidge1996}.

The absorption towards J0238$+$1636 (also known as AO 0235+164) is also seen at a restframe wavelength of 21~cm revealing a complex system of multiple absorbing clouds \citep{Roberts1976}. Two galaxies --J0238A and J0238A1 -- with small impact parameters ($1.1$\arcsec $=7$~kpc  and $1.9$\arcsec $=12$~kpc) between the galaxy position and the quasar sight line towards J0238$+$1636 were identified at the absorber redshift using [OII]-narrow-band imaging by \cite{Yanny1989}. {\it HST} WFPC2 imaging suggests that these are two compact components embedded in an extended nebula \citep{Chen2005}. 
The CO detections from these galaxies are presented for the first time in this study.
For consistency with previous publications we refer to these galaxies as J0238A and J0238A1. 

For the absorber towards J04230--0120 \citet{Klitsch2018} identified a total of four galaxies at the absorber redshift using the MUSE IFU. One of these four galaxies, J0423B, is also detected in CO(2--1) and CO(3--2)  emission. It is proposed that the gas seen in absorption traces either intra-group gas, or an outflow from J0423B.

The properties of the ALMA-detected galaxies are given in Table~\ref{TabGalEssential}. Flux maps and spectra are shown in Fig.~\ref{FigSpecandMaps}. The flux maps are integrated over $w_{20}$ which is also marked by the shaded area in the respective spectra. The spectra are measured using an aperture that encompasses the $3\sigma$ contours in the moment map. The line flux is determined by integrating over $w_{20}$. We use these multiple line detections to study the CO spectral line energy distribution (CO SLEDs) of these absorption-selected galaxies (see Fig.~\ref{FigCOSLED}).

We cannot use the Band 6 observations to measure the continuum flux of J0238A and J0238A1, because of strong residuals from the quasar continuum. In Band 7 we do not see any residuals, however, we also do not detect the two galaxies. The $3 \sigma$ upper limit assuming that the galaxies are not resolved is 0.1 mJy. 
We use a set of template spectra for starburst galaxies from \citet{Polletta2007} which we scale to the observed $870 \:\mu \text{m}$ flux converted to the rest wavelength. The fitted spectrum is then integrated in the wavelength range from $8$~-~$1000 \:\mu \text{m}$ to obtain the total far-infrared luminosity $L_{\rm FIR}$. This is converted to the SFR using the \citet{Kennicutt1998} relation. Based on this we derive the SFR limits reported in Table~\ref{TabPhysProp}.

\section{Analysis}

\begin{figure*}
\begin{minipage}{0.65\linewidth}
\begin{center}
\large{J0238+1636} \large{($z_{\rm abs} = 0.524$)}\vspace{-0.3cm}\\
\end{center}
\includegraphics[width = 0.45\linewidth]{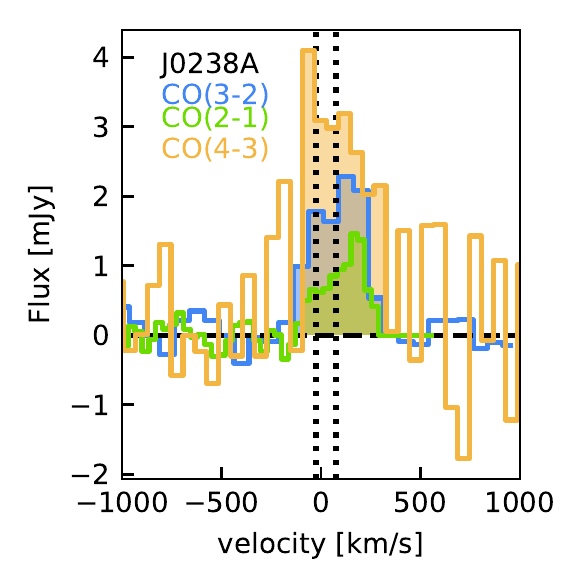}
\includegraphics[width = 0.45\linewidth]{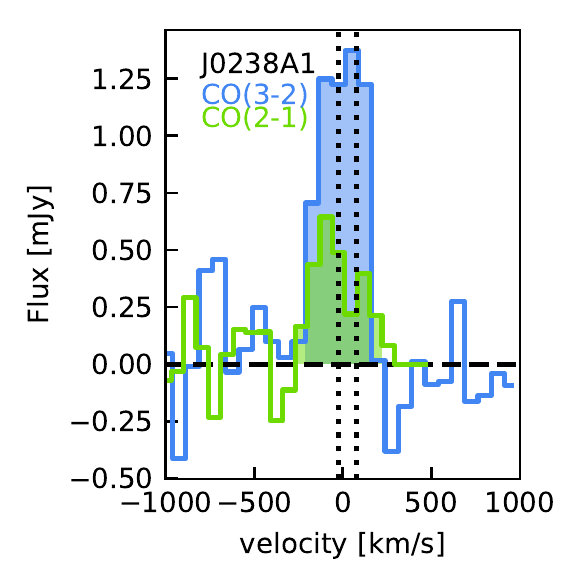}\vspace{-0.3cm}\\
\vspace{-0.3cm}
\includegraphics[width = 0.45\linewidth]{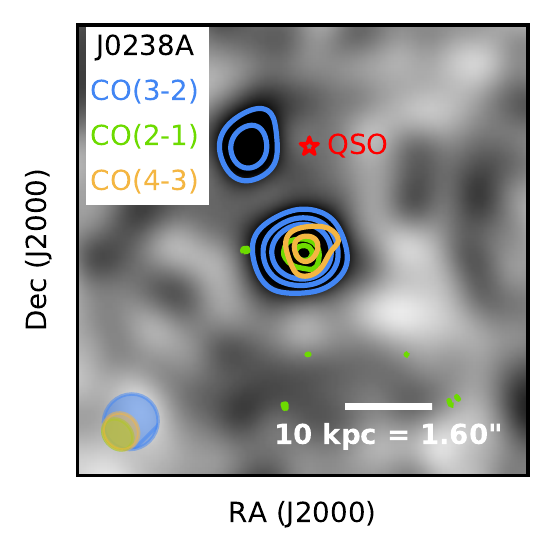}
\includegraphics[width = 0.45\linewidth]{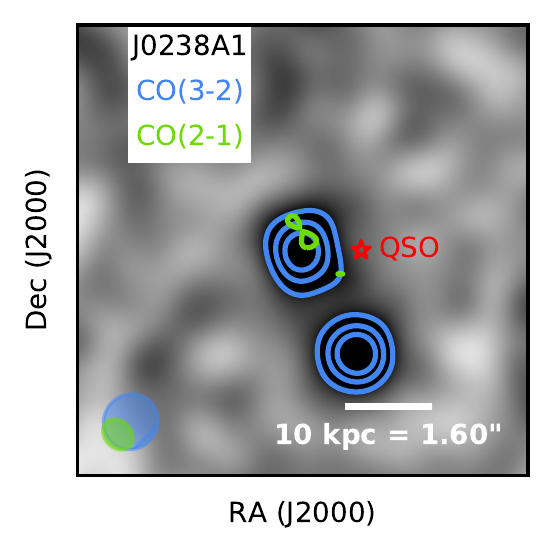}
\end{minipage}
\vline
\hfill
\begin{minipage}{0.33\linewidth}
\begin{center}
\large{J0423-0120} \large{($z_{\rm abs} = 0.633$)}\vspace{-0.3cm}\\
\end{center}
\includegraphics[width = 0.9\linewidth]{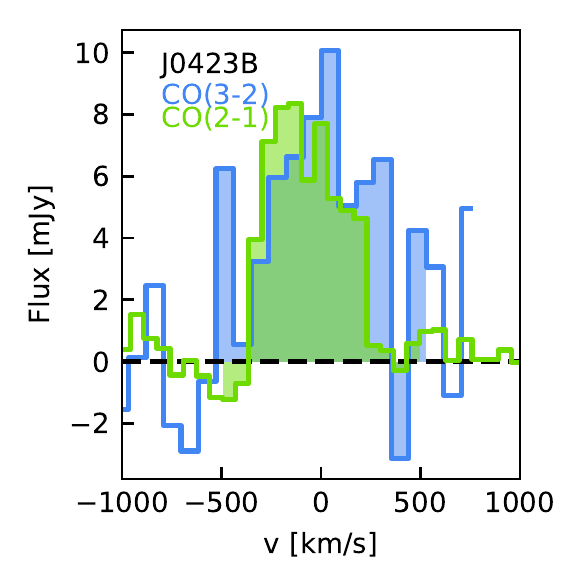} \vspace{-0.3cm} \\ \vspace{-0.3cm}
\includegraphics[width = 0.9\linewidth, trim = 180 0 200 0, clip]{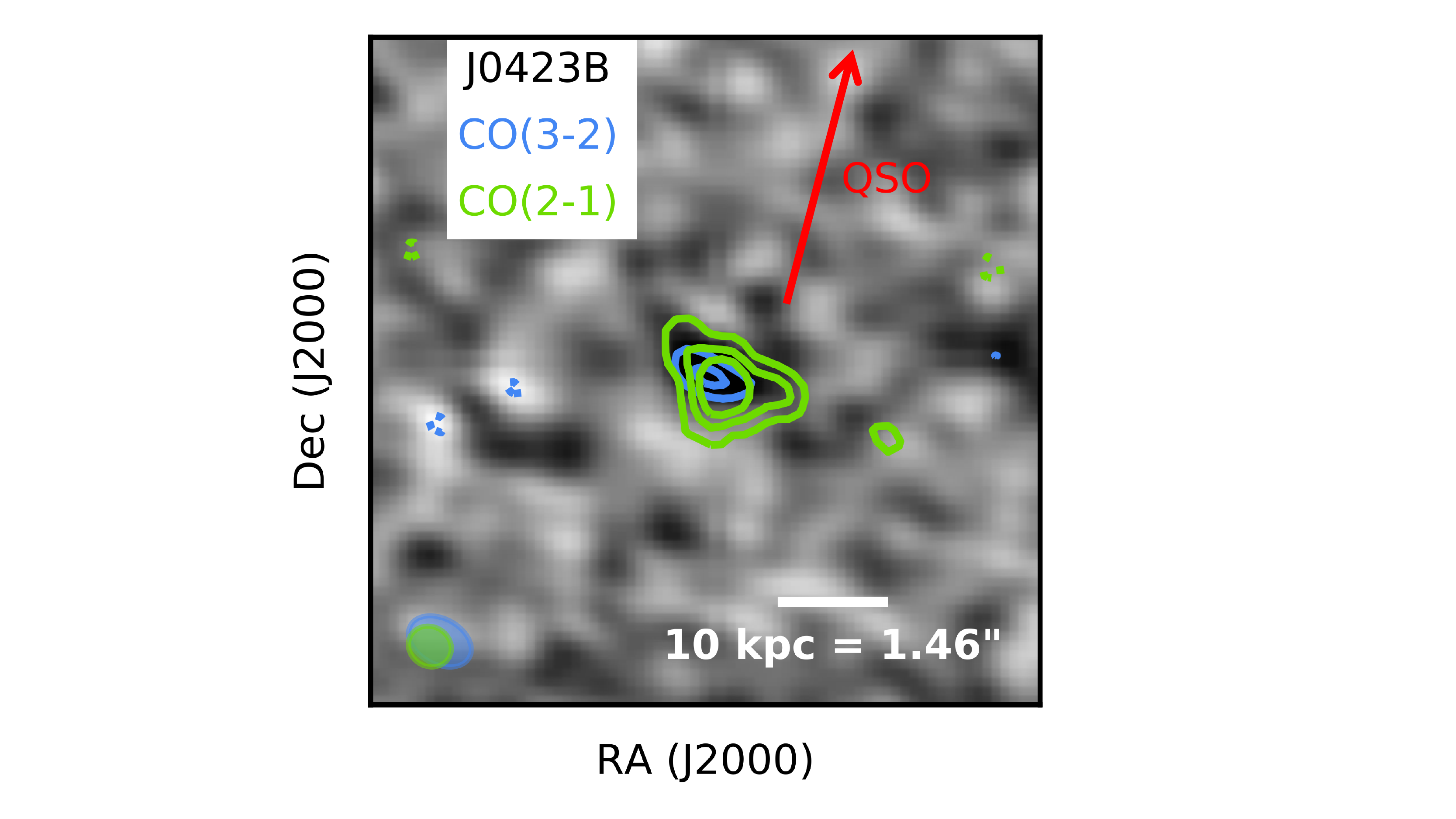}
\end{minipage}

\caption{Spectra (\textit{top}, primary beam corrected) and integrated flux maps (\textit{bottom}, not primary beam corrected) of the absorption-selected galaxies with multiple ALMACAL CO lines detected from the same galaxy. The figure for J0423B is reproduced from \citet{Klitsch2018}. \textit{Top:} A velocity of $0 \; \text{km s}^{-1}$ corresponds to the redshift of the main absorption component. The dotted lines in the spectra of J0238A and J0238A1 represent the position of the second and third most prominent features in the 21 cm absorption spectrum.
\textit{Bottom:} The CO(3--2) emission line integrated flux maps are shown in grey-scale for all three galaxies, the contours mark the $3\sigma$, $5\sigma$ and $7\sigma$ levels of the respective maps. For each map we integrate over the coloured region in the spectra shown above. The dotted contours mark the $-3\sigma$ levels. The sight-line towards the quasar is marked by a red star. These are the first detections of multiple CO emission lines from absorption-selected galaxies enabling us to constrain the energetics of their ISM.}
\label{FigSpecandMaps}
\end{figure*}

\begin{figure}
\includegraphics[width = 0.9\linewidth, trim = 0 5 0 5, clip]{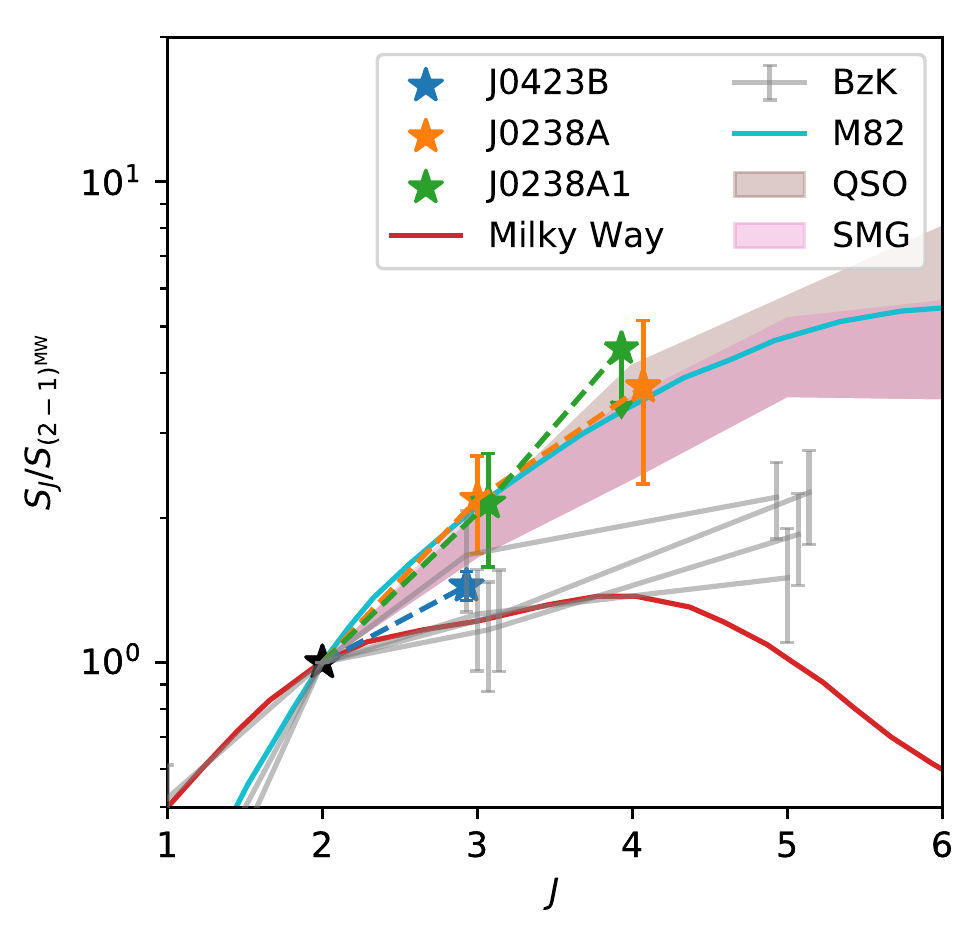}
\caption{CO Spectral line energy distribution of the three absorption-selected galaxies with multiple CO emission lines detected in ALMACAL. We show the CO SLED for $BzK$ galaxies \citep{Daddi2015}, the Milky Way, QSOs, SMGs and M82 \citep{Ivison2011, Carilli2013} for comparison. The CO SLEDs of two absorption-selected galaxies differ from the CO SLED of normal star forming galaxies at the low $J$ transitions. This suggests that the temperature and density of the molecular gas in these galaxies is higher than in normal star-forming galaxies and is more similar to high redshift SMGs or AGN. We find hints that using the Galactic line luminosity conversion factors to derive molecular gas masses might overestimate the molecular gas masses for some absorption-selected galaxies.}
\label{FigCOSLED}
\end{figure}

\begin{table*}
\caption{The derived physical properties of the CO-detected galaxies associated with two absorbers from our ALMA observations.}
\label{TabPhysProp}
\begin{tabular}{l c c c c c l}
\hline
Name & $\log$ N(HI) & $\log M_{\rm mol}$ [$\rm{M}_{\sun}/\alpha_{\rm CO}$] & $\log M_{\star}$ [$\rm{M}_{\sun}$] & SFR[M$_{\sun}$~yr $^{-1}$] & Z$_{\rm abs}$[Z$_{\sun}$] & type\\
\hline
J0238A & $21.70^{+0.08}_{-0.10}$ & $ 9.12 \pm 0.08$ & -- & $<50$ & $0.72 \pm 0.24$ & BAL QSO$^2$\\
J0238A1 & $21.70^{+0.08}_{-0.10}$ & $ 8.9 \pm 0.1$ & -- & $<50$ & $0.72 \pm 0.24$ & AGN $+$ starburst$^3$\\
J0423B & $18.54^{+0.07}_{-0.10}$ & $10.33 \pm 0.02$ & $11.2 \pm 0.1$ & $50 \pm 10$ & $>0.07$ & starburst$^1$\\
\hline
\end{tabular}
\begin{minipage}{\linewidth}
References: $^1$~\citet{Klitsch2018}, $^2$~\citet{Burbidge1996}, $^3$~\citet{Chen2005}
\end{minipage}
\end{table*}

At the redshift of the absorber towards J0238+1636, we find two CO-emitting galaxies. 
The absorption system is likely probing the joint gas distribution of both galaxies that was also seen in emission by \citet{Chen2005}.
 \citet{Roberts1976} reported a complex 21~cm absorption spectrum that might be due to several absorbing clouds. The main features are over-plotted in Fig.~\ref{FigSpecandMaps} and coincide with J0238A and J0238A1 in velocity space. 

We use the CO SLEDs shown in Fig.~\ref{FigCOSLED} as a diagnostic plot to distinguish between gas conditions similar to normal star-forming galaxies and more active systems. We note that the CO SLEDs are normalized to the \mbox{CO(2--1)} flux because the CO(1--0) emission line was not observed towards the absorption-selected galaxies. Differences in the CO SLEDs at higher $J$ transitions will appear smaller due to this normalization. 

The shape of the CO SLED determines the conversion of higher $J$ transition line fluxes to the CO(1--0) line flux, which is used to calculate the total molecular gas mass. The CO SLEDs of J0238A and J0238A1 are clearly distinct from the CO SLED of the Milky Way or $BzK$ galaxies. This suggests that the temperature and density in these absorption-selected galaxies are higher than in ``normal'' star-forming galaxies. 
J0423B at $z = 0.633$ on the other hand shows a CO SLED only slightly steeper than that of the Milky Way and comparable with that of ``normal'' star-forming galaxies at higher redshift ($z \sim1.5$). The SFR of this galaxy is $50\pm10 \; {\rm M}_{\sun} \; {\rm yr}^{-1}$ \citep{Klitsch2018} explaining the elevated CO SLED compared to the Milky Way. Hence, the Galactic conversion factor, $\alpha_{\rm CO} \simeq 4.6\: \rm{M}_{\sun}\: (\rm{K\: km\: s^{-1} pc^2})^{-1}$, and the Galactic CO line ratios might not be applicable. Instead, we argue that for the absorption-selected galaxies studied here, LIRG-like conversion factors are more appropriate. We convert the CO line flux to $L'_{\rm line}$ following the description from \citet{Solomon1992}.

To derive the molecular gas mass we convert the luminosity of the CO(2--1) transition to the CO(1--0) line luminosity. 
We convert the line luminosities using the median conversion factor $r_{21} = L'_{\rm CO(2-1)}/L'_{\rm CO(1-0)} = 0.9$ for LIRGs \citep{Papadopoulos2012a}. 
The resulting molecular gas masses are reported in Table~\ref{TabPhysProp} and are \mbox{$\sim 10^9 - 10^{10} {\rm M}_{\sun}/\alpha_{\rm CO}$.}

Comparing the L$'_{\rm CO}$--L$_{{\rm FIR}}$ relation we find that J0423B is consistent with local LIRGs and that J0238A and J0238A1 are in the lower envelope of local LIRGs \citep{Greve2014}.

\section{Discussion and Conclusion}

\begin{figure*}
\includegraphics[width = 0.45\linewidth]{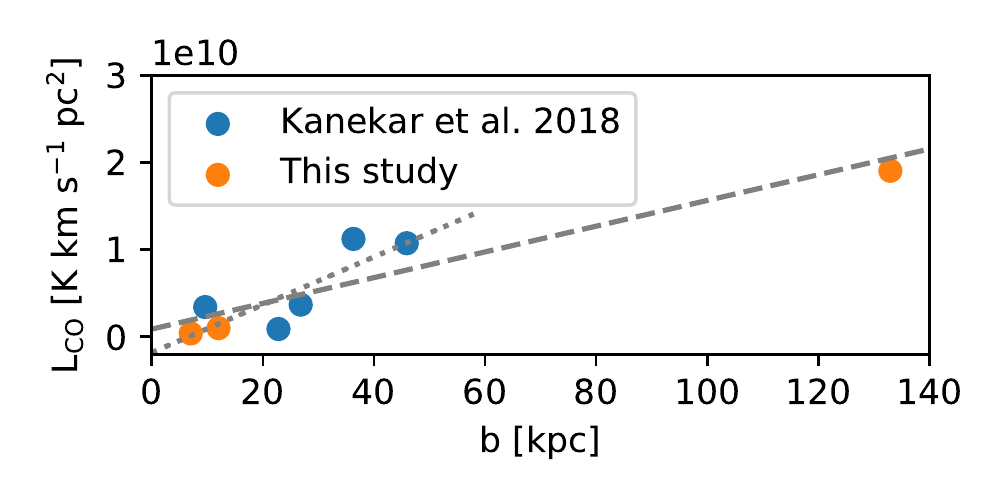}
\includegraphics[width = 0.45\linewidth]{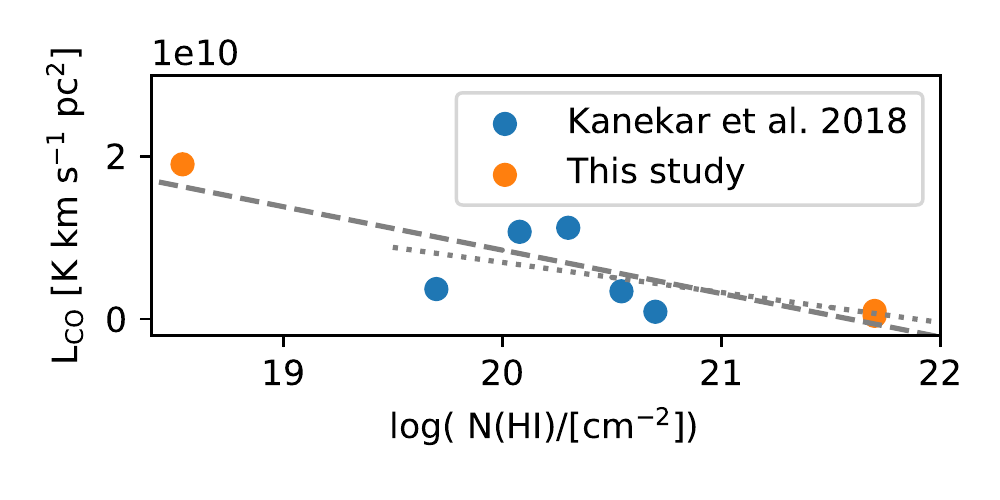}
\vspace{-0.5cm}
\caption{{\it Left:} CO(2-1) luminosity as a function of impact parameter. We show the absorption-selected galaxies presented by \citet{Kanekar2018}, and the galaxies presented here. {\it Right:} CO(2-1) luminosity as a function of H{\sc i} absorption column density. Errorbars are too small to be visible on these plots. We find correlation of $L_{\rm CO}$ with the impact parameter towards the QSO sight line and an anti-correlation of $L_{\rm CO}$ with the column density of the absorber. A fit of a linear function to the data is shown by the grey dashed line. We find the following best fits including J0423B: $ L_{\rm CO} [\text{K km s}^{-1} \text{pc}^2]=  1.5 \times 10^{8} \: b/ [\text{kpc}] + 9.0 \times 10^{8}$ and $ L_{\rm CO} [\text{K km s}^{-1} \text{pc}^2] = - 5.3 \times 10^9 \: \log({\rm N(HI)/[\text{cm}^{-2}]}) + 1.2 \times 10^{11} $ and excluding J0423B:  $ L_{\rm CO} [\text{K km s}^{-1} \text{pc}^2] =  2.7 \times 10^{8} \: b / [\text{kpc}] - 1.8 \times 10^{9}$ and $L_{\rm CO} [\text{K km s}^{-1} \text{pc}^2] = - 3.7 \times 10^9 \: \log({\rm N(HI)/[\text{cm}^{-2}]}) + 8.1 \times 10^{10} $. }
\label{FigKPC18Comp}
\end{figure*}

We report the detection of multiple CO emission lines from absorption-selected galaxies discovered in our ALMACAL survey. 
The CO SLEDs for two galaxies are more excited than those of "normal" star-forming galaxies indicating that the molecular gas temperature and density are higher in these galaxies. This confirms the optical identification as BAL QSO and mix of AGN and starburst, respectively.
These findings suggest that the Galactic CO line ratios and $\alpha_{\rm CO}$ conversion factor might not be applicable to derive molecular gas masses of all damped Ly$\alpha$ galaxies. \\
\indent We derive molecular gas masses of $10^{8.9}$ to $10^{10.33}$ ${\rm M}_{\sun}/\alpha_{\rm CO}$ using a $r_{21}$ conversion factor of 0.9. 
These derived molecular gas masses are smaller than those reported by \citet{Kanekar2018}, who used a Galactic $r_{21}$ ratio and a Galactic CO-to-H$_2$ conversion factor. Their resulting molecular gas masses are $ 10^{9.8 - 10.9} \times (\alpha_{\rm CO}/4.36) \times (0.63/r_{21}) {\rm M}_{\sun}$. As discussed above for a fraction of these galaxies a smaller CO-to-H$_2$ conversion factor could be more appropriate. Direct evidence for a more excited ISM from other wavelength regimes such as the detection of an AGN in J0238A and A1 should always be taken into account when choosing appropriate line ratios and $\alpha_{\rm CO}$. In fact, at least one galaxy in \citet{Kanekar2018} is reported to interact with another galaxy \citep{Straka2016}. \\
\indent Comparing the line fluxes, we find that for J0423B the CO(2--1) flux is higher than in any of the galaxies presented by \citet{Kanekar2018} while for J0238A and A1 the fluxes are on the low end of the ones reported by \citet{Kanekar2018}. 
This is also illustrated in Fig.~\ref{FigKPC18Comp}, where we show the CO(2--1) luminosity of the absorber host galaxies as a function of  impact parameters and column densities of the absorbers from this study and \citet{Kanekar2018}. Using the Pearson correlation coefficient we find a correlation between $L_{\rm CO}$ and the impact parameter with a p-value of 0.01 and an anti-correlation between $L_{\rm CO}$ and log N(H{\sc i}) with a p-value of 0.002. To test whether these relations are only driven by J0423B, we exclude it from the analysis resulting in p-values of 0.1 and 0.01. In these two samples low impact parameter absorbers probe the high density CGM close to low mass galaxies, while at high impact parameters the lower density diffuse gas around high mass galaxies is traced. Using a significance level of 0.05 we would only reject the null-hypothesis of a correlation between $L_{\rm CO}$ and the impact parameter. However, the results are based on a small dataset and more data is needed to further investigate these correlations.\\
\indent The question remains whether the three galaxies presented here are special among absorption selected systems? \citet{Verdes-Montenegro2001} proposed a scenario in which -- depending on the evolutionary phase of the group -- up to $\sim60$\% of the H{\sc i} gas is stripped from the galaxies by tidal interactions and resides in a diffuse intra-group medium. This was tested in the Hickson groups. 
\citet{Borthakur2010} reported a diffuse gas fraction of $\sim20 - 50$\% increasing the cross-section of H{\sc i} compared to isolated galaxies. This would result in a higher likelihood to probe interacting galaxies with Ly$\alpha$ absorbers. Such interacting systems are also more likely to contain galaxies with enhanced star formation and AGN activity. Hence, it could be expected that the CO SLEDs of some absorption-selected galaxies are more excited than typical gas rich field galaxies.  
To test further this hypothesis more data are needed to populate the CO SLEDs of absorption-selected galaxies. In general it is important to study the CO SLED to properly derive molecular gas masses. Similarly, galactic winds can increase the cross-section of H{\sc i}, where the most energetic events will produce the highest sky-coverage. Such galaxies will deviate from the Galactic CO SLED as well.\\
\indent Since both interactions and winds are predicted to be enhanced at higher redshift this might have an important impact the types of galaxies identified through surveys of absorption-selected galaxies in the distant universe.

\section*{Acknowledgements}

The authors would like to thank Helmut Dannerbauer for an interesting discussion on CO SLEDs.
AK acknowledges support from the STFC grant ST/P000541/1 and Durham University. 
CP is grateful to the ESO and the DFG cluster of excellence 'Origin and Structure of the Universe' for support.
IRS acknowledges support from the ERC Advanced Grant DUSTYGAL (321334), a Royal Society/Wolfson Merit Award and STFC (ST/P000541/1).
ALMA is a partnership of ESO (representing its member states), NSF (USA) and NINS (Japan), together with NRC (Canada), MOST and ASIAA (Taiwan), and KASI (Republic of Korea), in cooperation with the Republic of Chile. The Joint ALMA Observatory is operated by ESO, AUI/NRAO and NAOJ. This paper makes use of the following ALMA data: 
ADS/JAO.ALMA\#2013.1.00989.S, \#2016.1.01435.S, \#2015.1.01084.S, \#2015.1.00419.S, \#2015.1.00598.S, \#2016.1.01553.S, \#2015.1.00187.S, \#2016.1.00683.S, \#2016.1.01010.S, \#2015.1.00932.S, \#2015.1.00723.S, \#2016.1.01262.S, \#2016.1.00505.S, \#2016.1.01184.S, \#2015.1.01144.S, \#2015.1.00330.S, \#2012.1.00090.S, \#2012.1.00979.S, \#2012.1.00245.S, \#2013.1.00001.S, \#2013.1.00534.S, \#2013.1.00989.S, \#2013.1.01225.S, \#2015.1.00024.S, \#2015.1.00456.S, \#2015.1.01290.S, \#2015.1.01503.S, \#2016.1.00434.S, \#2016.1.01305.S.
This research made use of Astropy, a community-developed core Python package for Astronomy \citep{Astropy2018}.




\bibliographystyle{mnras}

\begin{thebibliography}{}
\makeatletter
\relax
\def\mn@urlcharsother{\let\do\@makeother \do\$\do\&\do\#\do\^\do\_\do\%\do\~}
\def\mn@doi{\begingroup\mn@urlcharsother \@ifnextchar [ {\mn@doi@}
  {\mn@doi@[]}}
\def\mn@doi@[#1]#2{\def\@tempa{#1}\ifx\@tempa\@empty \href
  {http://dx.doi.org/#2} {doi:#2}\else \href {http://dx.doi.org/#2} {#1}\fi
  \endgroup}
\def\mn@eprint#1#2{\mn@eprint@#1:#2::\@nil}
\def\mn@eprint@arXiv#1{\href {http://arxiv.org/abs/#1} {{\tt arXiv:#1}}}
\def\mn@eprint@dblp#1{\href {http://dblp.uni-trier.de/rec/bibtex/#1.xml}
  {dblp:#1}}
\def\mn@eprint@#1:#2:#3:#4\@nil{\def\@tempa {#1}\def\@tempb {#2}\def\@tempc
  {#3}\ifx \@tempc \@empty \let \@tempc \@tempb \let \@tempb \@tempa \fi \ifx
  \@tempb \@empty \def\@tempb {arXiv}\fi \@ifundefined
  {mn@eprint@\@tempb}{\@tempb:\@tempc}{\expandafter \expandafter \csname
  mn@eprint@\@tempb\endcsname \expandafter{\@tempc}}}

\bibitem[\protect\citeauthoryear{{Bielby} et~al.,}{{Bielby}
  et~al.}{2016}]{Bielby2016}
{Bielby} R.~M.,  et~al., 2016, \mn@doi [\mnras] {10.1093/mnras/stv2914}, \href
  {https://ui.adsabs.harvard.edu/#abs/2016MNRAS.456.4061B} {456, 4061}

\bibitem[\protect\citeauthoryear{{Bird} \& {et al.}}{{Bird} \& {et
  al.}}{2014}]{Bird2014}
{Bird} S.,  {et al.} 2014, \mn@doi [\mnras] {10.1093/mnras/stu1923}, \href
  {https://ui.adsabs.harvard.edu/#abs/2014MNRAS.445.2313B} {445, 2313}

\bibitem[\protect\citeauthoryear{{Borthakur} \& {et al.}}{{Borthakur} \& {et
  al.}}{2010}]{Borthakur2010}
{Borthakur} S.,  {et al.} 2010, \mn@doi [\apj] {10.1088/0004-637X/710/1/385},
  \href {https://ui.adsabs.harvard.edu/#abs/2010ApJ...710..385B} {710, 385}

\bibitem[\protect\citeauthoryear{{Burbidge} et~al.}{{Burbidge}
  et~al.}{1996}]{Burbidge1996}
{Burbidge} E.~M.,  et~al., 1996, \mn@doi [\aj] {10.1086/118199}, \href
  {https://ui.adsabs.harvard.edu/#abs/1996AJ....112.2533B} {112, 2533}

\bibitem[\protect\citeauthoryear{{Carilli} \& {Walter}}{{Carilli} \&
  {Walter}}{2013}]{Carilli2013}
{Carilli} C.~L.,  {Walter} F.,  2013, \mn@doi [ARA\&A]
  {10.1146/annurev-astro-082812-140953}, \href
  {https://ui.adsabs.harvard.edu/#abs/2013ARA&A..51..105C} {51, 105}

\bibitem[\protect\citeauthoryear{{Chen} \& {et al.}}{{Chen} \& {et
  al.}}{2005}]{Chen2005}
{Chen} H.-W.,  {et al.} 2005, \mn@doi [\apj] {10.1086/427088}, \href
  {https://ui.adsabs.harvard.edu/#abs/2005ApJ...620..703C} {620, 703}

\bibitem[\protect\citeauthoryear{{Daddi} et~al.,}{{Daddi}
  et~al.}{2015}]{Daddi2015}
{Daddi} E.,  et~al., 2015, \mn@doi [\aap] {10.1051/0004-6361/201425043}, \href
  {https://ui.adsabs.harvard.edu/#abs/2015A&A...577A..46D} {577, A46}

\bibitem[\protect\citeauthoryear{{Fumagalli} et~al.}{{Fumagalli}
  et~al.}{2010}]{Fumagalli2010}
{Fumagalli} M.,  et~al., 2010, \mn@doi [\mnras]
  {10.1111/j.1365-2966.2010.17113.x}, \href
  {http://adsabs.harvard.edu/abs/2010MNRAS.408..362F} {408, 362}

\bibitem[\protect\citeauthoryear{{Fumagalli} et~al.}{{Fumagalli}
  et~al.}{2011}]{Fumagalli2011}
{Fumagalli} M.,  et~al., 2011, \mn@doi [\mnras]
  {10.1111/j.1365-2966.2011.19599.x}, \href
  {https://ui.adsabs.harvard.edu/#abs/2011MNRAS.418.1796F} {418, 1796}

\bibitem[\protect\citeauthoryear{{Fumagalli} et~al.,}{{Fumagalli}
  et~al.}{2017}]{Fumagalli2017}
{Fumagalli} M.,  et~al., 2017, \mn@doi [\mnras] {10.1093/mnras/stx1896}, \href
  {https://ui.adsabs.harvard.edu/#abs/2017MNRAS.471.3686F} {471, 3686}

\bibitem[\protect\citeauthoryear{{Fynbo} et~al.,}{{Fynbo}
  et~al.}{2018}]{Fynbo2018}
{Fynbo} J.~P.~U.,  et~al., 2018, \mn@doi [\mnras] {10.1093/mnras/sty1520},
  \href {http://adsabs.harvard.edu/abs/2018MNRAS.479.2126F} {479, 2126}

\bibitem[\protect\citeauthoryear{{Greve} et~al.,}{{Greve}
  et~al.}{2014}]{Greve2014}
{Greve} T.~R.,  et~al., 2014, \mn@doi [\apj] {10.1088/0004-637X/794/2/142},
  \href {http://adsabs.harvard.edu/abs/2014ApJ...794..142G} {794, 142}

\bibitem[\protect\citeauthoryear{{Ivison} \& {et al.}}{{Ivison} \& {et
  al.}}{2011}]{Ivison2011}
{Ivison} R.~J.,  {et al.} 2011, \mn@doi [\mnras]
  {10.1111/j.1365-2966.2010.18028.x}, \href
  {https://ui.adsabs.harvard.edu/#abs/2011MNRAS.412.1913I} {412, 1913}

\bibitem[\protect\citeauthoryear{{Junkkarinen}, {Cohen}, {Beaver}, {Burbidge},
  {Lyons}  \& {Madejski}}{{Junkkarinen} et~al.}{2004}]{Junkkarinen2004}
{Junkkarinen} V.~T.,  {Cohen} R.~D.,  {Beaver} E.~A.,  {Burbidge} E.~M.,
  {Lyons} R.~W.,   {Madejski} G.,  2004, \mn@doi [\apj] {10.1086/423777}, \href
  {https://ui.adsabs.harvard.edu/#abs/2004ApJ...614..658J} {614, 658}

\bibitem[\protect\citeauthoryear{{Kanekar} et~al.,}{{Kanekar}
  et~al.}{2018}]{Kanekar2018}
{Kanekar} N.,  et~al., 2018, \mn@doi [\apj] {10.3847/2041-8213/aab6ab}, \href
  {https://ui.adsabs.harvard.edu/#abs/2018ApJ...856L..23K} {856, L23}

\bibitem[\protect\citeauthoryear{{Kennicutt}}{{Kennicutt}}{1998}]{Kennicutt1998}
{Kennicutt} Jr. R.~C.,  1998, \mn@doi [\araa] {10.1146/annurev.astro.36.1.189},
  \href {http://adsabs.harvard.edu/abs/1998ARA%26A..36..189K} {36, 189}

\bibitem[\protect\citeauthoryear{{Klitsch}, {P{\'e}roux}, {Zwaan}, {Smail},
  {Oteo}, {Biggs}, {Popping}  \& {Swinbank}}{{Klitsch}
  et~al.}{2018}]{Klitsch2018}
{Klitsch} A.,  {P{\'e}roux} C.,  {Zwaan} M.~A.,  {Smail} I.,  {Oteo} I.,
  {Biggs} A.~D.,  {Popping} G.,   {Swinbank} A.~M.,  2018, \mn@doi [\mnras]
  {10.1093/mnras/stx3184}, \href
  {https://ui.adsabs.harvard.edu/#abs/2018MNRAS.475..492K} {475, 492}

\bibitem[\protect\citeauthoryear{{Krogager} et~al.}{{Krogager}
  et~al.}{2017}]{Krogager2017}
{Krogager} J.~K.,  et~al., 2017, \mn@doi [\mnras] {10.1093/mnras/stx1011},
  \href {https://ui.adsabs.harvard.edu/#abs/2017MNRAS.469.2959K} {469, 2959}

\bibitem[\protect\citeauthoryear{{M{\o}ller} et~al.,}{{M{\o}ller}
  et~al.}{2018}]{Mller2018}
{M{\o}ller} P.,  et~al., 2018, \mn@doi [\mnras] {10.1093/mnras/stx2845}, \href
  {https://ui.adsabs.harvard.edu/#abs/2018MNRAS.474.4039M} {474, 4039}

\bibitem[\protect\citeauthoryear{{Muzahid}, {Srianand}  \&
  {Charlton}}{{Muzahid} et~al.}{2015}]{Muzahid2015}
{Muzahid} S.,  {Srianand} R.,   {Charlton} J.,  2015, \mn@doi [\mnras]
  {10.1093/mnras/stv133}, \href
  {https://ui.adsabs.harvard.edu/#abs/2015MNRAS.448.2840M} {448, 2840}

\bibitem[\protect\citeauthoryear{{Neeleman} \& {et al.}}{{Neeleman} \& {et
  al.}}{2018}]{Neeleman2018}
{Neeleman} M.,  {et al.} 2018, \mn@doi [\apjl] {10.3847/2041-8213/aab5b1},
  \href {http://adsabs.harvard.edu/abs/2018ApJ...856L..12N} {856, L12}

\bibitem[\protect\citeauthoryear{{Neeleman} et~al.,}{{Neeleman}
  et~al.}{2016}]{Neeleman2016}
{Neeleman} M.,  et~al., 2016, \mn@doi [\apj] {10.3847/2041-8205/820/2/L39},
  \href {https://ui.adsabs.harvard.edu/#abs/2016ApJ...820L..39N} {820, L39}

\bibitem[\protect\citeauthoryear{{Noterdaeme} \& {et al.}}{{Noterdaeme} \& {et
  al.}}{2008}]{Noterdaeme2008}
{Noterdaeme} P.,  {et al.} 2008, \mn@doi [\aap] {10.1051/0004-6361:20078780},
  \href {https://ui.adsabs.harvard.edu/#abs/2008A&A...481..327N} {481, 327}

\bibitem[\protect\citeauthoryear{{Oteo} \& {et al.}}{{Oteo} \& {et
  al.}}{2016}]{Oteo2016}
{Oteo} I.,  {et al.} 2016, \mn@doi [\apj] {10.3847/0004-637X/822/1/36}, \href
  {https://ui.adsabs.harvard.edu/#abs/2016ApJ...822...36O} {822, 36}

\bibitem[\protect\citeauthoryear{{Papadopoulos} et~al.}{{Papadopoulos}
  et~al.}{2012}]{Papadopoulos2012a}
{Papadopoulos} P.~P.,  et~al., 2012, \mn@doi [\mnras]
  {10.1111/j.1365-2966.2012.21001.x}, \href
  {https://ui.adsabs.harvard.edu/#abs/2012MNRAS.426.2601P} {426, 2601}

\bibitem[\protect\citeauthoryear{{P{\'e}roux} et~al.}{{P{\'e}roux}
  et~al.}{2011}]{Peroux2011}
{P{\'e}roux} C.,  et~al., 2011, \mn@doi [\mnras]
  {10.1111/j.1365-2966.2010.17598.x}, \href
  {https://ui.adsabs.harvard.edu/#abs/2011MNRAS.410.2237P} {410, 2237}

\bibitem[\protect\citeauthoryear{{P{\'e}roux} et~al.,}{{P{\'e}roux}
  et~al.}{2017}]{Peroux2017}
{P{\'e}roux} C.,  et~al., 2017, \mn@doi [\mnras] {10.1093/mnras/stw2444}, \href
  {https://ui.adsabs.harvard.edu/#abs/2017MNRAS.464.2053P} {464, 2053}

\bibitem[\protect\citeauthoryear{{Polletta} et~al.,}{{Polletta}
  et~al.}{2007}]{Polletta2007}
{Polletta} M.,  et~al., 2007, \mn@doi [\apj] {10.1086/518113}, \href
  {http://adsabs.harvard.edu/abs/2007ApJ...663...81P} {663, 81}

\bibitem[\protect\citeauthoryear{{Rahmani} et~al.,}{{Rahmani}
  et~al.}{2017}]{Rahmani2017}
{Rahmani} H.,  et~al., 2017, preprint, \href
  {http://adsabs.harvard.edu/abs/2017arXiv171008398R} {} (\mn@eprint {arXiv}
  {1710.08398})

\bibitem[\protect\citeauthoryear{{Rao}, {Turnshek}  \& {Nestor}}{{Rao}
  et~al.}{2006}]{Rao2006}
{Rao} S.~M.,  {Turnshek} D.~A.,   {Nestor} D.~B.,  2006, \mn@doi [\apj]
  {10.1086/498132}, \href
  {https://ui.adsabs.harvard.edu/#abs/2006ApJ...636..610R} {636, 610}

\bibitem[\protect\citeauthoryear{{Roberts} et~al.}{{Roberts}
  et~al.}{1976}]{Roberts1976}
{Roberts} M.~S.,  et~al., 1976, \mn@doi [\aj] {10.1086/111885}, \href
  {https://ui.adsabs.harvard.edu/#abs/1976AJ.....81..293R} {81, 293}

\bibitem[\protect\citeauthoryear{{Rudie}, {Newman}  \& {Murphy}}{{Rudie}
  et~al.}{2017}]{Rudie2017}
{Rudie} G.~C.,  {Newman} A.~B.,   {Murphy} M.~T.,  2017, \mn@doi [\apj]
  {10.3847/1538-4357/aa74d7}, \href
  {https://ui.adsabs.harvard.edu/#abs/2017ApJ...843...98R} {843, 98}

\bibitem[\protect\citeauthoryear{{Solomon}, {Downes}  \& {Radford}}{{Solomon}
  et~al.}{1992}]{Solomon1992}
{Solomon} P.~M.,  {Downes} D.,   {Radford} S.~J.~E.,  1992, \mn@doi [\apj]
  {10.1086/186304}, \href
  {https://ui.adsabs.harvard.edu/#abs/1992ApJ...387L..55S} {387, L55}

\bibitem[\protect\citeauthoryear{{Straka} \& {et al.}}{{Straka} \& {et
  al.}}{2016}]{Straka2016}
{Straka} L.~A.,  {et al.} 2016, \mn@doi [\mnras] {10.1093/mnras/stw508}, \href
  {http://adsabs.harvard.edu/abs/2016MNRAS.458.3760S} {458, 3760}

\bibitem[\protect\citeauthoryear{{The Astropy Collaboration} et~al.,}{{The
  Astropy Collaboration} et~al.}{2018}]{Astropy2018}
{The Astropy Collaboration} et~al., 2018, preprint, \href
  {http://adsabs.harvard.edu/abs/2018arXiv180102634T} {} (\mn@eprint {arXiv}
  {1801.02634})

\bibitem[\protect\citeauthoryear{{Verdes-Montenegro} \& {et
  al.}}{{Verdes-Montenegro} \& {et al.}}{2001}]{Verdes-Montenegro2001}
{Verdes-Montenegro} L.,  {et al.} 2001, \mn@doi [\aap]
  {10.1051/0004-6361:20011127}, \href
  {https://ui.adsabs.harvard.edu/#abs/2001A&A...377..812V} {377, 812}

\bibitem[\protect\citeauthoryear{{Wolfe} et~al.}{{Wolfe}
  et~al.}{1986}]{Wolfe1986}
{Wolfe} A.~M.,  et~al., 1986, \mn@doi [The Astrophysical Journal Supplement
  Series] {10.1086/191114}, \href
  {https://ui.adsabs.harvard.edu/#abs/1986ApJS...61..249W} {61, 249}

\bibitem[\protect\citeauthoryear{{Yanny}, {York}  \& {Gallagher}}{{Yanny}
  et~al.}{1989}]{Yanny1989}
{Yanny} B.,  {York} D.~G.,   {Gallagher} J.~S.,  1989, \mn@doi [\apj]
  {10.1086/167231}, \href
  {https://ui.adsabs.harvard.edu/#abs/1989ApJ...338..735Y} {338, 735}

\bibitem[\protect\citeauthoryear{{Zwaan} \& {Prochaska}}{{Zwaan} \&
  {Prochaska}}{2006}]{Zwaan2006}
{Zwaan} M.~A.,  {Prochaska} J.~X.,  2006, \mn@doi [\apj] {10.1086/503191},
  \href {https://ui.adsabs.harvard.edu/#abs/2006ApJ...643..675Z} {643, 675}

\bibitem[\protect\citeauthoryear{{van de Voort} \& {Schaye}}{{van de Voort} \&
  {Schaye}}{2012}]{vandeVoort2012}
{van de Voort} F.,  {Schaye} J.,  2012, \mn@doi [\mnras]
  {10.1111/j.1365-2966.2012.20949.x}, \href
  {https://ui.adsabs.harvard.edu/#abs/2012MNRAS.423.2991V} {423, 2991}

\makeatother
\end{thebibliography}








\bsp	
\label{lastpage}
\end{document}